\documentclass[aps,prb,twocolumn,showpacs,superscriptaddress,floatfix]{revtex4}
\usepackage{ulem}
\usepackage{graphicx}% Include figure files
\usepackage{dcolumn}% Align table columns on decimal point
\usepackage{bm}% bold math

\usepackage{cancel}

\begin{document}

\title{Linear response of doped graphene sheets to vector potentials}
\author{A. Principi}
\affiliation{NEST-CNR-INFM and Dipartimento di Fisica dell'Universit\`a di Pisa, I-56127 Pisa, Italy}
\author{Marco Polini}
\email{m.polini@sns.it}
\homepage{http://qti.sns.it}
\affiliation{NEST-CNR-INFM and Scuola Normale Superiore, I-56126 Pisa, Italy}
\author{G. Vignale}
\homepage{http://web.missouri.edu/~vignaleg/web_site/}
\affiliation{Department of Physics and Astronomy, University of Missouri, Columbia, Missouri 65211, USA}

\begin{abstract}
A two-dimensional gas of massless Dirac fermions (MDFs) is a very useful model to describe low-energy electrons in monolayer graphene. Because the MDF current operator is directly proportional to the (sublattice) pseudospin operator, the MDF current-current response function, which describes the response to a vector potential, happens to coincide with the pseudospin-pseudospin response function.
In this work we present analytical results for the wavevector- and frequency-dependent longitudinal and transverse pseudospin-pseudospin response functions of noninteracting MDFs. The transverse response in the static limit is then used to calculate the noninteracting orbital magnetic susceptibility. These results are a starting point for the construction of approximate pseudospin-pseudospin response functions that would take into account electron-electron interactions (for example at the random-phase-approximation level). They also constitute a very useful input for future applications of current-density-functional theory to graphene sheets subjected to time- and spatially-varying vector potentials. 
\end{abstract}


\maketitle

\section{Introduction}
\label{sect:intro}

Graphene, a monolayer of Carbon atoms packed in a 2D honeycomb lattice, is a recently realized ambipolar gapless semiconductor 
that has attracted enormous interest~\cite{geim_nat_mat_2007,katsnelson_ssc_2007,geim_pt_2007,castro_neto_rmp_2009}. Electrons in graphene are described at low energies by a spin-independent massless Dirac Hamiltonian, which ultimately originates from the non-Bravais nature of the 2D honeycomb lattice. 
The two inequivalent sites in the unit cell of this lattice are analogous to the two spin orientations of a spin-$1/2$ particle along the $+{\hat {\bm z}}$ and $-{\hat {\bm z}}$ directions (the ${\hat {\bm z}}$ axis being perpendicular to the graphene plane).  This observation opens the way to an elegant description of electrons in graphene as particles endowed with a {\it pseudospin} degree-of-freedom~\cite{geim_nat_mat_2007,katsnelson_ssc_2007,geim_pt_2007,castro_neto_rmp_2009} (in addition to the regular spin and valley degrees of freedom which play a passive role here). This quantum degree-of-freedom has a number of very intriguing implications on the electronic properties of this material, most of which have been reviewed in Refs.~\onlinecite{geim_nat_mat_2007,katsnelson_ssc_2007,geim_pt_2007,castro_neto_rmp_2009}.

Graphene offers a unique example of a new paradigm in condensed matter physics: a truly 2D non-Galileian invariant electron liquid (see for example Refs.~\onlinecite{jiang_prl_2007,deacon_prb_2007,iyengar_prb_2007,mueller_prb_2008,polini_pseudospin_2009}). 
This non-Galileian invariant nature of graphene is linked to non-trivial many-body renormalizations of 
various properties of doped graphene sheets~\cite{polini_pseudospin_2009}, such as the plasmon dispersion relation and the optical conductivity~\cite{infrared_spectroscopy}, even at very long-wavelengths.  Both these properties are 
controlled by the linear response function of 2D MDFs to a vector potential, {\it i.e.} by the current-current response function.
Because in the case of MDFs a vector potential couples to the pseudospin-density-fluctuation operator, this response function happens to coincide, apart from a trivial proportionality factor, with the pseudospin-pseudospin linear-response function.

In this work we present analytical results for the wavevector- and frequency-dependent longitudinal and transverse pseudospin-pseudospin response functions of {\it noninteracting} 2D MDFs.

Our paper is organized as follows.  In Section~\ref{sect:model} we present the model and some basic definitions. 
In Section~\ref{sect:undoped} we present analytical results for the wavevector- and frequency-dependent longitudinal and transverse pseudospin-pseudospin response functions of a noninteracting {\it undoped} 2D MDF system.  In Section~\ref{sect:doped} we report corresponding results for the {\it doped} system. In Sect.~\ref{sect:diamagnetic_suscept} we use the analytical results on the transverse response to calculate the orbital magnetic susceptibility of noninteracting MDFs. 
Finally, in Section~\ref{sect:summary} we present a brief summary of our main results and mention their usefulness 
for the construction of (i) response functions that would take into account electron-electron interactions, and (ii) exchange-correlation functionals needed in applications of current-density-functional theory~\cite{vignale_prl_1987} to graphene sheets in the presence of inhomogeneous vector potentials. Finally, two appendices report cumbersome technical details and calculations.

\section{The model and basic definitions}
\label{sect:model}

Graphene's honeycomb lattice has two-atoms per unit 
cell and its $\pi$-valence band and $\pi^*$-conduction band touch at two inequivalent 
points, $K$ and $K'$, in the honeycomb lattice Brillouin-zone. The energy bands near {\it e.g.} 
the $K$ point are described at low energies by a spin-independent massless Dirac Hamiltonian ($\hbar=1$)
\begin{equation}\label{eq:Dirac}
{\hat {\cal H}}_{\rm D}= 
v \sum_{{\bm k}, \alpha, \beta} {\hat \psi}^\dagger_{{\bm k}, \alpha} ({\bm \sigma}_{\alpha\beta} \cdot {\bm k}) {\hat \psi}_{{\bm k}, \beta}~,
\end{equation}
where ${\hat \psi}^\dagger_{{\bm k}, \alpha}$ (${\hat \psi}_{{\bm k}, \alpha}$) creates (destroys) an electron with momentum ${\bm k}$ and sublattice pseudospin $\alpha$ and ${\bm \sigma}=(\sigma^x,\sigma^y)$ is a vector constructed 
with two Pauli matrices $\{\sigma^i,i=x,y\}$, which act on the sublattice pseudospin degree-of-freedom.  
Because we are interested in the linear-response functions to smoothly-varying vector potentials, to which different spins and valleys contribute 
independently, in Eq.~(\ref{eq:Dirac}) we have retained only sublattice degrees of freedom. As emphasized repeatedly in the literature (see {\it e.g.} Refs.~\onlinecite{castro_neto_rmp_2009,polini_pseudospin_2009,barlas_prl_2007}), because of the presence of an infinite sea of negative-energy states, the MDF model (\ref{eq:Dirac}) must be accompanied by an ultraviolet cut-off for the wavevector integrals, $k_{\rm max}$. 

The response to a vector potential ${\bm A}({\bm q}, \omega)$ is controlled by the current-current linear-response function
$\chi_{{\bm j}{\bm j}}({\bm q},\omega)$, which is defined by the usual Kubo product~\cite{Giuliani_and_Vignale} 
\begin{eqnarray}
\chi_{AB}(\omega) &=& \frac{1}{S}\langle\langle {\hat A}; {\hat B}\rangle\rangle_\omega \nonumber\\
&\equiv&
-\frac{i}{S}\lim_{\epsilon \to 0^+} \int_0^{\infty}dt \langle [{\hat A}(t), {\hat B}(0)]\rangle e^{i\omega t}e^{-\epsilon t}~.
\end{eqnarray}
The current-density operator ${\hat {\bm j}}_{\bm q}$ can be easily found from the continuity equation. The density operator ${\hat \rho}_{\bm q}$ corresponding to the Hamiltonian ${\hat {\cal H}}_{\rm D}$ reads
\begin{equation}
{\hat \rho}_{\bm q} =\sum_{{\bm k},\alpha} {\hat \psi}^\dagger_{{\bm k}-{\bm q}, \alpha} {\hat \psi}_{{\bm k}, \alpha}~,
\end{equation}
and obeys the usual continuity equation 
\begin{equation}\label{eq:continuity}
i\partial_t {\hat \rho}_{\bm q} = [{\hat \rho}_{\bm q}, {\hat {\cal H}}_{\rm D}] = {\bm q} \cdot {\hat {\bm j}}_{\bm q}~,
\end{equation}
with the current-density operator ${\hat {\bm j}}_{\bm q}$ that has a rather unusual form~\cite{katsnelson_ssc_2007},
\begin{equation}\label{eq:current_pseudospin_relation}
{\hat {\bm j}}_{\bm q} = v \sum_{{\bm k}, \alpha, \beta} {\hat \psi}^\dagger_{{\bm k}-{\bm q}, \alpha} {\bm \sigma}_{\alpha\beta} {\hat \psi}_{{\bm k}, \beta}~.
\end{equation}
The current-density operator for MDFs is proportional to the pseudospin-density operator. 
Due to Eq.~(\ref{eq:current_pseudospin_relation}) thus, the current-current response function 
$\chi_{{\bm j}{\bm j}}({\bm q},\omega)$ is proportional to the pseudospin-pseudospin response function, 
$\chi_{{\bm j}{\bm j}}({\bm q},\omega) = v^2 \chi_{{\bm \sigma}{\bm \sigma}}({\bm q},\omega)$. 

In what follows we will calculate the noninteracting {\it longitudinal} pseudospin-pseudospin response function, {\it i.e.} $\chi^{(0)}_{\sigma_x\sigma_x} (q{\hat {\bm x}}, \omega)$ with ${\bm q}$ oriented along the ${\hat {\bm x}}$ direction, ${\bm q}= q {\hat {\bm x}}$, and the {\it transverse} response function, {\it i.e.} $\chi^{(0)}_{\sigma_x\sigma_x} (q{\hat {\bm y}}, \omega)$ with ${\bm q}$ oriented along the ${\hat {\bm y}}$ direction, ${\bm q}= q {\hat {\bm y}}$.
These are given by~\cite{Giuliani_and_Vignale} (per spin and per valley)
\begin{eqnarray}\label{eq:lindhard}
\chi^{(0)}_{\sigma_x\sigma_x} ({\bm q}, \omega) &=& \frac{1}{S}\lim_{\epsilon \to 0^+}\sum_{\bm k}\sum_{\lambda, \lambda'} \frac{n^{(0)}_{{\bm k}, \lambda} - n^{(0)}_{{\bm k}+{\bm q}, \lambda'}}{\omega + \varepsilon_{{\bm k}, \lambda'} - \varepsilon_{{\bm k}+{\bm q}, \lambda} + i\epsilon}\nonumber\\
&\times& |\langle\chi_{\lambda}({\bm k})|\sigma_x|\chi_{\lambda'}({\bm k}+{\bm q})\rangle|^2~,
\end{eqnarray}
where $S$ is the area of the system, $\varepsilon_{{\bm k}, \lambda}= \lambda v k$ are single-particle Dirac energies, 
$n^{(0)}_{{\bm k}, \lambda}$ are noninteracting band-occupation factors and 
\begin{equation}
\chi_\lambda({\bm k})=\frac{1}{\sqrt{2}}\left(
\begin{array}{c}
1\\
\lambda e^{i\varphi_{\bm k}}
\end{array}
\right)
\end{equation}
are two-component pseudospinors. 
Here $\lambda = + 1$ labels the conduction band and $\lambda = - 1$ the valence band and $\varphi_{\bm k}$ is the angle between ${\bm k}$ and the ${\hat {\bm x}}$ axis, which physically 
denotes the momentum-dependent phase difference 
between wavefunction amplitudes on the A and B sublattices of graphene's honeycomb lattice. 
The matrix-element factor on the second line of Eq.~(\ref{eq:lindhard}) is given by
\begin{equation}\label{eq:sum_angles}
|\langle\chi_{\lambda}({\bm k})|\sigma_x|\chi_{\lambda'}({\bm k}+{\bm q})\rangle|^2= \frac{1+\lambda\lambda' \cos(\varphi_{\bm k}+\varphi_{{\bm k}+{\bm q}})}{2}~.
\end{equation}

Note that, because of the continuity equation (\ref{eq:continuity}), the {\it longitudinal} pseudospin-pseudospin response function 
is related to the density-density response function $\chi_{\rho\rho}$. In our case this relations reads
\begin{equation}\label{eq:eom_mdf}
\chi_{\rho\rho}(q,\omega) =
\frac{vq}{\omega^2}  \langle [{\hat \sigma}^x_{\bm q}, {\hat \rho}_{-{\bm q}}]\rangle
+\frac{v^2q^2}{\omega^2}\chi_{\sigma^x \sigma^x}(q{\hat {\bm x}},\omega)~.
\end{equation}
This formula holds only at finite $\omega$.
The first term on the r.h.s. of Eq.~(\ref{eq:eom_mdf}), which is an anomalous commutator 
because of the presence of the infinite sea of negative energy states, is {\it purely real} and must be handled with great care~\cite{sabio_prb_2008,polini_pseudospin_2009}. As discussed in Ref.~\onlinecite{polini_pseudospin_2009}, it is easy to show that in the noninteracting limit
\begin{equation}\label{eq:anomalous_comm}
\frac{1}{vq}\langle [{\hat \sigma}^x_{\bm q}, {\hat \rho}_{-{\bm q}}]\rangle \to \frac{\varepsilon_{\rm max}}{4\pi v^2}~,
\end{equation}
where $\varepsilon_{\rm max}/v = k_{\rm max}$ is an ultraviolet wavevector cut-off. Eq.~(\ref{eq:anomalous_comm}) is valid to leading order in the limit $k_{\rm max} \to \infty$.

\section{Undoped case}
\label{sect:undoped}

We first calculate the response functions $\chi^{(0{\rm u})}_{\sigma_x\sigma_x}$ of the undoped system, {\it i.e.} the system in which the Fermi energy lies at the Dirac point. In this case only the band $\varepsilon_{{\bm k},-}$ is full with an occupation factor $n^{(0)}_{{\bm k}, -}=1$, while the upper band $\varepsilon_{{\bm k},+}$ is empty (all necessary technical details are summarized in Appendix~\ref{app:undoped}).

In the longitudinal channel, for what stated at the end of Sect.~\ref{sect:model}, we find the following relation: 
\begin{equation}\label{eq:im_long_undoped}
\Im m~\chi^{(0{\rm u})}_{\sigma_x\sigma_x}(q{\hat {\bm x}},\omega) = \frac{\omega^2}{v^2 q^2}\Im m~\chi^{(0{\rm u})}_{\rho\rho}(q,\omega)
\end{equation}
where
\begin{equation}
\Im m~\chi^{(0{\rm  u})}_{\rho\rho}(q,\omega)  = -{\rm sgn}(\omega)\frac{q^2}{16}\frac{\Theta(\omega^2-v^2q^2)}{\sqrt{\omega^2-v^2q^2}}
\end{equation}
is the imaginary part of the well-known density-density response function of noninteracting MDFs at half filling~\cite{old_works}. Using the Kramers-Kr\"onig dispersion relations one immediately finds the result for the real part $\Re e~\chi^{(0{\rm u})}_{\sigma_x\sigma_x}(q{\hat {\bm x}},\omega)$:
\begin{equation}\label{eq:re_long_undoped}
\Re e~\chi^{(0{\rm  u})}_{\sigma_x\sigma_x} (q{\hat {\bm x}}, \omega) = -\frac{\varepsilon_{\rm max}}{4\pi v^2}+ \frac{\omega^2}{v^2 q^2}\Re e~\chi^{(0{\rm u})}_{\rho\rho}(q,\omega)~,
\end{equation}
where~\cite{old_works}
\begin{equation}
\Re e~\chi^{(0{\rm  u})}_{\rho\rho}(q,\omega)  = -\frac{q^2}{16}\frac{\Theta(v^2q^2-\omega^2)}{\sqrt{v^2q^2-\omega^2}}~.
\end{equation}

In the trasverse channel, starting again from Eq.~(\ref{eq:lindhard}) but with ${\bm q} =q {\hat {\bm y}}$, we find
\begin{equation}\label{eq:im_trans_undoped}
\Im m~\chi^{(0{\rm u})}_{\sigma_x \sigma_x} (q{\hat {\bm y}}, \omega) = -{\rm sgn}(\omega)\frac{\Theta(\omega^2 - v^2 q^2)}{16 v^2} \sqrt{\omega^2 - v^2q^2}~.
\end{equation}
Note that the previous equation can also be written as
\begin{eqnarray}\label{eq:im_trans_undoped_cute}
\Im m~\chi^{(0{\rm u})}_{\sigma_x \sigma_x} (q{\hat {\bm y}}, \omega) = \frac{\omega^2 - v^2 q^2}{v^2 q^2}\Im m~\chi^{(0{\rm u})}_{\rho\rho}(q,\omega)~.
\end{eqnarray}
This relation reminds us of Eq.~(\ref{eq:im_long_undoped}), the only difference being that in the transverse case 
we have the factor $(\omega^2 - v^2 q^2)/(v^2 q^2)$ rather than the ``gauge-invariance factor" $\omega^2/(v^2 q^2)$. This simple relation between the transverse pseudospin-pseudospin response function and the density-density response function, however, holds only in the undoped case.

Finally, for the real part of the transverse response function, again using the Kramers-Kr\"onig relations, we find
\begin{equation}\label{eq:re_trans_undoped}
\Re e~\chi^{(0{\rm  u})}_{\sigma_x\sigma_x} (q{\hat {\bm y}}, \omega) = -\frac{\varepsilon_{\rm max}}{4\pi v^2} + \frac{\omega^2-v^2q^2}{v^2 q^2}\Re e~\chi^{(0{\rm u})}_{\rho\rho}(q,\omega)~.
\end{equation}

In summary, we find that the longitudinal undoped pseudospin responses of the model described by (\ref{eq:Dirac}) with the ultraviolet cut-off $k_{\rm max}$ are given by
\begin{widetext}
\begin{equation}\label{eq:long}
\left\{
\begin{array}{ll}
{\displaystyle \Re e~\chi^{(0{\rm  u})}_{\sigma_x\sigma_x} (q{\hat {\bm x}}, \omega)} & = {\displaystyle -\frac{\varepsilon_{\rm max}}{4\pi v^2} -\frac{\omega^2}{16 v^2}\frac{\Theta(v^2q^2-\omega^2)}{\sqrt{v^2q^2-\omega^2}}} \vspace{0.2 cm}
\\
{\displaystyle \Im m~\chi^{(0{\rm  u})}_{\sigma_x\sigma_x} (q{\hat {\bm x}}, \omega)} &= {\displaystyle -{\rm sgn}(\omega)\frac{\omega^2}{16 v^2}\frac{\Theta(\omega^2-v^2q^2)}{\sqrt{\omega^2-v^2q^2}}}
\end{array}
\right.~,
\end{equation}
while the transverse undoped responses are given by
\begin{equation}\label{eq:trans}
\left\{
\begin{array}{ll}
{\displaystyle \Re e~\chi^{(0{\rm  u})}_{\sigma_x\sigma_x} (q{\hat {\bm y}}, \omega)} &= {\displaystyle -\frac{\varepsilon_{\rm max}}{4\pi v^2} + \frac{\sqrt{v^2q^2- \omega^2}}{16 v^2}\Theta(v^2q^2-\omega^2)} \vspace{0.2 cm} \\
{\displaystyle \Im m~\chi^{(0{\rm  u})}_{\sigma_x\sigma_x} (q{\hat {\bm y}}, \omega)} &= {\displaystyle -{\rm sgn}(\omega)
\frac{\sqrt{\omega^2 - v^2 q^2}}{16 v^2} \Theta(\omega^2-v^2q^2)}
\end{array}
\right.~.
\end{equation}
\end{widetext}
Eqs.~(\ref{eq:long}) and~(\ref{eq:trans}) constitute the first novel results of this work. Note that longitudinal and transverse responses coincide at $q=0$ and that the anomalous terms in the first lines of Eqs.~(\ref{eq:long}) and~(\ref{eq:trans}) [first terms on the r.h.s. of both equations] are identical. 

Two crucial remarks are in order at this point. Because of gauge invariance, the real physical system cannot respond to a {\it static} longitudinal vector potential. More precisely, the longitudinal current-current response function of the physical system should vanish for $\omega=0$ and every $q$, while the transverse one should vanish for $\omega=0$ and $q \to 0$. After a quick look at Eqs.~(\ref{eq:long}) and~(\ref{eq:trans}) one can easily see that this is not what happens to the response functions of the model system (\ref{eq:Dirac}), simply because of the presence of the cut-off dependent constant term $-\varepsilon_{\rm max}/(4\pi v^2)$. As discussed in Ref.~\onlinecite{polini_pseudospin_2009} (see footnote 33), this unphysical finite response is due to the fact that rigorous gauge-invariance of the model described by ${\hat {\cal H}}_{\rm D}$ is broken by the ultraviolet cut-off $k_{\rm max}$. Thus, the static response functions of the model system have to be corrected {\it ad hoc} in order to restore gauge invariance: the quantity $-\varepsilon_{\rm max}/(4\pi v^2)$ has to be subtracted away~\cite{polini_pseudospin_2009} from the first lines in Eqs.~(\ref{eq:long}) and~(\ref{eq:trans}). On the other hand, we would like to remark that in the limit $q=0$ and for finite $\omega$ the constant term $-\varepsilon_{\rm max}/(4\pi v^2)$ is not only {\it physical} ({\it i.e.} it describes the response of the $\lambda = -1$ valence band to a uniform vector potential in the regime in which repopulation of states is not allowed~\cite{polini_pseudospin_2009}) but also {\it necessary}: when Eq.~(\ref{eq:long}) is substituted inside Eq.~(\ref{eq:eom_mdf}), this term indeed cancels exactly the anomalous commutator [by virtue of Eq.~(\ref{eq:anomalous_comm})], giving a density-density response function which is independent of $k_{\rm max}$, as it should~\cite{old_works}.

The pseudospin response functions can be written in a more compact form using the following complex functions
\begin{equation}\label{eq:long_F}
F_{\rm L}(q,\omega) = \frac{\omega^2}{16 \pi v^2}\frac{1}{\sqrt{\omega^2 - v^2 q^2}}
\end{equation}
and
\begin{equation}\label{eq:trans_F}
F_{\rm T}(q,\omega) = \frac{\omega^2-v^2q^2}{\omega^2} F_{\rm L}(q,\omega)~.
\end{equation}
We find
\begin{equation}
\chi^{(0{\rm  u})}_{\sigma_x\sigma_x} (q{\hat {\bm x}}({\hat {\bm y}}), \omega) = -\frac{\varepsilon_{\rm max}}{4\pi v^2} - i \pi F_{{\rm L}({\rm T})}(q,\omega)~.
\end{equation}

Finally, note that the real part of the long-wavelength longitudinal conductivity $\sigma(\omega)$ of the undoped system is given by
\begin{equation}\label{eq:universal}
\Re e~\sigma(\omega) = 
 -\frac{v^2 e^2}{\omega}~\lim_{q\to 0} \Im m~\chi^{(0{\rm  u})}_{\sigma_x\sigma_x} (q{\hat {\bm x}}, \omega) = \frac{e^2}{16}~.
\end{equation}
Restoring $\hbar$ and introducing the $g_{\rm s}g_{\rm v} =4$ spin-valley degeneracy we find the usual ``universal" frequency-independent value~\cite{infrared_spectroscopy} $\Re e~\sigma(\omega)  = e^2/(4 \hbar)$.

\section{Doped case}
\label{sect:doped}

\begin{figure}[t]
\begin{center}
\includegraphics[width=1.00\linewidth]{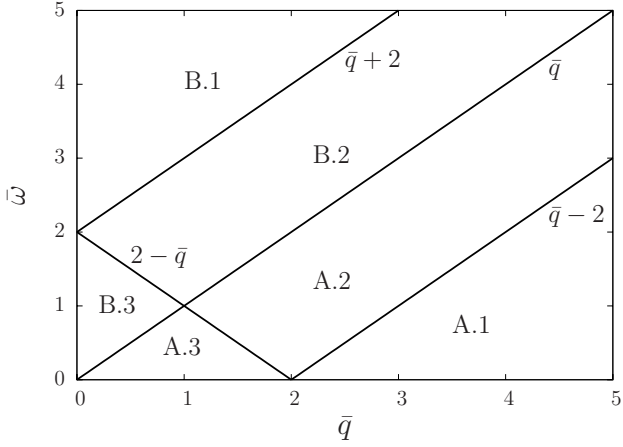}
\end{center}
\caption{Different regions for the behavior of $\chi^{(0)}_{\sigma_x\sigma_x}({\bm q},\omega)$. 
In this figure we have introduced dimensionless variables: ${\bar q} = q/ k_{\rm F}$ and ${\bar \omega} = \omega/ \varepsilon_{\rm F}$
The regions ${\rm B}.1$ and ${\rm B}.2$ are characterized by a continuum of {\it interband} electron-hole pairs. The regions ${\rm A}.2$ and ${\rm A}.3$ are characterized by a continuum of {\it intraband} electron-hole pairs. $\Im m\chi^{(0)}_{\sigma_x\sigma_x}({\bm q}, \omega)=0$ in regions ${\rm A}.1$ and ${\rm B}.3$.\label{fig:one}}
\end{figure}

We now pass to calculate the response functions $\chi^{(0)}_{\sigma_x\sigma_x}$ of the doped system, {\it i.e.} the system in which the Fermi energy lies, for example, above the Dirac point. In this case the band $\varepsilon_{{\bm k},-}$ is completely full with the usual 
occupation factor $n^{(0)}_{{\bm k}, -}=1$, while the upper band $\varepsilon_{{\bm k},+}$ is filled only up to the Fermi energy $\varepsilon_{\rm F} =v k_{\rm F}$, where $k_{\rm F} =\sqrt{4 \pi n}$. In the case of finite doping, we find more convenient to evaluate Eq.~(\ref{eq:lindhard}) on the imaginary-frequency axis ({\it i.e.} by letting $\omega \rightarrow i\omega$) and then to perform a standard analytical continuation to the real-frequency axis (more details are given in Appendix~\ref{app:doped}).

\subsection{Longitudinal channel}
\label{sect:long_doped}

Even though, as discussed before, the longitudinal response function is determined by the density-density response function~\cite{barlas_prl_2007,wunsch_njp_2006,hwang_prb_2007} {\it via} the continuity equation, in this Section we report, for the sake of completeness, expressions for $\chi^{(0)}_{\sigma_x\sigma_x}(q{\hat {\bm x}},\omega)$. 

The contribution to $\chi^{(0)}_{\sigma_x\sigma_x}(q{\hat {\bm x}},\omega)$ due to doping, 
$\delta \chi^{(0)}_{\sigma_x\sigma_x}(q{\hat {\bm x}},\omega)$, is introduced according to
\begin{equation}
\chi^{(0)}_{\sigma_x\sigma_x}(q{\hat {\bm x}},\omega) = \chi^{(0{\rm u})}_{\sigma_x\sigma_x}(q{\hat {\bm x}},\omega) + \delta \chi^{(0)}_{\sigma_x\sigma_x}(q{\hat {\bm x}},\omega)~.
\end{equation}
Explicit expressions for the real and imaginary parts of $\chi^{(0{\rm u})}_{\sigma_x\sigma_x}(q{\hat {\bm x}},\omega)$ have been given before in Sect.~\ref{sect:undoped}. Below we thus report only expressions for the doping-dependent quantity $\delta \chi^{(0)}_{\sigma_x\sigma_x}(q{\hat {\bm x}},\omega)$. 

Following Wunsch {\it et al.}~\cite{wunsch_njp_2006} we introduce the complex function
\begin{equation}
G_{\rm L} (z) = z\sqrt{z^2 - 1} - \ln(z + \sqrt{z^2 - 1})
\end{equation}
and
\begin{equation}
\omega_\pm = \frac{2\varepsilon_{\rm F} \pm \omega}{vq}~.
\end{equation}
Using $G_{\rm L}(z)$ and the function $F_{\rm L}(q,\omega)$ introduced above in Eq.~(\ref{eq:trans_F}) we find
\begin{eqnarray}
\delta \chi^{(0)}_{\sigma_x\sigma_x}(q{\hat {\bm x}},\omega)&=& -\frac{\omega^2}{v^2 q^2}\frac{\varepsilon_{\rm F}}{2\pi v^2} + F_{\rm L}(q, \omega) \{
G_{\rm L}(\omega_+)  \nonumber \\ 
&-& \Theta(\omega_- - 1) [G_{\rm L}(\omega_-) - i \pi] \nonumber \\
&-& \Theta(1 - \omega_-) G_{\rm L}(-\omega_-)\}~.
\end{eqnarray}
We now provide more explicit expressions for $\delta \chi^{(0)}_{\sigma_x\sigma_x}(q{\hat {\bm x}},\omega)$ in terms of real functions. With reference to Fig.~\ref{fig:one} we find the following analytical results.

\subsubsection{Region A.1}
For 
%$\omega < vq$ and 
$\omega < vq - 2\varepsilon_{\rm F}$:
\begin{eqnarray}
\Re e~\delta \chi^{(0)}_{\sigma_x\sigma_x} (q{\hat {\bm x}}, \omega) &=&  
-\frac{\omega^2}{v^2 q^2}\frac{\varepsilon_{\rm F}}{2\pi v^2} \nonumber\\
&+& \frac{\omega^2}{16\pi v^2\sqrt{v^2 q^2 - \omega^2}} a_{\rm L}(q,\omega) \nonumber\\
\end{eqnarray}
and $\Im m~\delta \chi^{(0)}_{\sigma_x\sigma_x} (q\hat{\bm{x}}, \omega) = 0$. Here 
\begin{eqnarray}
a_{\rm L}(q,\omega) &=& \arcsin\left(\omega_+\right) + \omega_+\sqrt{1 -\omega_+^2}\nonumber\\
&+& \arcsin\left(\omega_-\right) \nonumber + \omega_-\sqrt{1 - \omega_-^2}~.
\end{eqnarray}

\subsubsection{Region A.2}
For $\omega < vq$ and $\omega >|2\varepsilon_{\rm F}- vq|$:
\begin{eqnarray}
\Re e~\delta \chi^{(0)}_{\sigma_x\sigma_x} (q\hat{\bm{x}}, \omega) &=& \frac{\omega^2}{16 v^2\sqrt{v^2 q^2 - \omega^2}} - \frac{\omega^2}{v^2 q^2}\frac{\varepsilon_{\rm F}}{2\pi v^2} \nonumber\\
&-& \frac{\omega^2}{16\pi v^2\sqrt{v^2 q^2 - \omega^2}} b_{\rm L}(q,\omega)
\end{eqnarray}
and
\begin{eqnarray}
\Im m~\delta \chi^{(0)}_{\sigma_x\sigma_x} (q\hat{\bm{x}}, \omega) = \frac{\omega^2}{16\pi v^2\sqrt{v^2 q^2 - \omega^2}} c_{\rm L}(q,\omega)~.
\end{eqnarray}
Here
\begin{equation}
b_{\rm L}(q,\omega) = \arccos(\omega_-) - \omega_-\sqrt{1 - \omega_-^2}
\end{equation}
and
\begin{equation}
c_{\rm L}(q,\omega) = \ln\left(\omega_+ + \sqrt{\omega_+^2 - 1}\right) - \omega_+\sqrt{\omega_+^2 - 1}~.
\end{equation}

\subsubsection{Region A.3}
For $\omega < vq$ and $\omega<2\varepsilon_{\rm F}-vq$:
\begin{equation}
\Re e~\delta \chi^{(0)}_{\sigma_x\sigma_x} (q\hat{\bm{x}}, \omega) = \frac{\omega^2}{16 v^2\sqrt{v^2 q^2 - \omega^2}} - \frac{\omega^2}{v^2 q^2}\frac{\varepsilon_{\rm F}}{2\pi v^2}
\end{equation}
and
\begin{equation}
\Im m~\delta \chi^{(0)}_{\sigma_x\sigma_x} (q\hat{\bm{x}}, \omega) = \frac{\omega^2}{16\pi v^2\sqrt{v^2 q^2 - \omega^2}} d_{\rm L}(q, \omega)~.
\end{equation}
Here
\begin{eqnarray}
d_{\rm L}(q,\omega) &=& \ln\left(\frac{\omega_+ + \sqrt{\omega_+^2 - 1}}{\omega_- + \sqrt{\omega_-^2 - 1}}\right) - \omega_+\sqrt{\omega_+^2 - 1} \nonumber\\
&+& \omega_-\sqrt{\omega_-^2 - 1}~.
\end{eqnarray}

\subsubsection{Region B.1}
For 
%$\omega > vq$ and 
$\omega > 2\varepsilon_{\rm F} + vq$:
\begin{eqnarray}
\Re e~\delta \chi^{(0)}_{\sigma_x\sigma_x} (q\hat{\bm{x}}, \omega) &=& -\frac{\omega^2}{v^2 q^2}\frac{\varepsilon_{\rm F}}{2\pi v^2} \nonumber\\
&-& \frac{\omega^2}{16\pi v^2\sqrt{\omega^2 - v^2 q^2}} e_{\rm L}(q,\omega) \nonumber\\
\end{eqnarray}
and $\Im m~\delta \chi^{(0)}_{\sigma_x\sigma_x} (q\hat{\bm{x}}, \omega) = 0$. Here
\begin{eqnarray}
e_{\rm L} (q,\omega) &=& \ln\left(\frac{\omega_+ + \sqrt{\omega_+^2 - 1}}{\sqrt{\omega_-^2 - 1}-\omega_-}\right) - \omega_+\sqrt{\omega_+^2 - 1} \nonumber\\
&-& \omega_-\sqrt{\omega_-^2 - 1}~.
\end{eqnarray}

\subsubsection{Region B.2}
For  $\omega > vq$, $\omega > 2\varepsilon_{\rm F}-vq$, and $\omega < 2\varepsilon_{\rm F} +vq$:
\begin{eqnarray}
\Re e~\delta \chi^{(0)}_{\sigma_x\sigma_x} (q\hat{\bm{x}}, \omega) &=& -\frac{\omega^2}{v^2 q^2}\frac{\varepsilon_{\rm F}}{2\pi v^2} \nonumber\\
&-& \frac{\omega^2}{16\pi v^2\sqrt{\omega^2 - v^2 q^2}} f_{\rm L}(q,\omega) \nonumber\\
\end{eqnarray}
and
\begin{eqnarray}
\Im m~\delta \chi^{(0)}_{\sigma_x\sigma_x} (q\hat{\bm{x}}, \omega) &=& \frac{\omega^2}{16 v^2\sqrt{\omega^2 - v^2 q^2}} \nonumber\\
&-& \frac{\omega^2}{16\pi v^2\sqrt{\omega^2 - v^2 q^2}} g_{\rm L}(q,\omega) ~. \nonumber\\
\end{eqnarray}
Here
\begin{equation}
f_{\rm L} (q,\omega) = \ln\left(\omega_+ + \sqrt{\omega_+^2 - 1}\right) - \omega_+\sqrt{\omega_+^2 - 1}
\end{equation}
and
\begin{equation}
g_{\rm L} (q,\omega) = \arccos\left(\omega_-\right) - \omega_-\sqrt{1 - \omega_-^2}~.
\end{equation}

\subsubsection{Region B.3}
For $\omega> vq$ and $\omega <2\varepsilon_{\rm F} - vq$:
\begin{eqnarray}\label{eq:re_plasmon_region}
\nonumber \Re e~\delta \chi^{(0)}_{\sigma_x\sigma_x} (q\hat{\bm{x}}, \omega) &=& -\frac{\omega^2}{v^2 q^2}\frac{\varepsilon_{\rm F}}{2\pi v^2} \nonumber\\
&-& \frac{\omega^2}{16\pi v^2\sqrt{\omega^2 - v^2 q^2}} h_{\rm L} (q,\omega) \nonumber\\
\end{eqnarray}
and
\begin{equation}\label{eq:im_plasmon_region}
\Im m~\delta \chi^{(0)}_{\sigma_x, \sigma_x} (q\hat{\bm{x}}, \omega) = \frac{\omega^2}{16 v^2\sqrt{\omega^2 - v^2 q^2}}~.
\end{equation}
Here
\begin{eqnarray}\label{eq:hL}
h_{\rm L} (q,\omega) &=& \ln\left(\frac{\omega_+ + \sqrt{\omega_+^2 - 1}}{\omega_- + \sqrt{\omega_-^2 - 1}}\right) - \omega_+\sqrt{\omega_+^2 - 1} \nonumber\\
&+&  \omega_-\sqrt{\omega_-^2 - 1}~.
\end{eqnarray}

We would now like to remind the reader that in the region B.3 of the $(q,\omega)$ plane the {\it interacting} system possesses a collective plasmon mode $\omega_{\rm pl} = \omega_{\rm pl}(q)$, which, within the simple random phase approximation (RPA)~\cite{Giuliani_and_Vignale,wunsch_njp_2006,hwang_prb_2007,polini_prb_2008}, can be found by solving the equation
\begin{equation}\label{eq:plasmon}
1 - v_q \Re e \chi^{(0)}_{\rho \rho}(q,\omega) = 0~,
\end{equation}
where $v_q = 2\pi e^2/(\epsilon q)$ is the 2D Fourier transform of the Coulomb potential ($\epsilon$ being an average dielectric constant that depends on the environment surrounding the graphene flake). Using Eq.~(\ref{eq:eom_mdf}) we can write Eq.~(\ref{eq:plasmon}) in the more appealing form
\begin{equation}\label{eq:plasmon_current_current}
1- v_q \frac{v^2 q^2}{\omega^2}  \left[\frac{\varepsilon_{\rm max}}{4 \pi v^2} + \Re e~\chi^{(0)}_{\sigma_x \sigma_x}(q{\hat {\bm x}},\omega)\right] = 0~.
\end{equation}
Using the microscopic expression for $\Re e~\chi^{(0u)}_{\sigma_x \sigma_x}(q{\hat {\bm x}},\omega)$ in Eq.~(\ref{eq:long}) and Eq.~(\ref{eq:re_plasmon_region}), expanding the expression in square brackets in Eq.~(\ref{eq:plasmon_current_current}) in powers of $q/k_{\rm F}$ up to fourth order,  introducing the $g_{\rm s} g_{\rm v} = 4$ spin-valley degeneracy factor, and restoring for a moment $\hbar$, we find that the plasmon dispersion is given by
\begin{equation}\label{eq:plasmon_dispersion}
\omega_{\rm pl}(q \to 0) = \sqrt{\frac{2 \pi n e^2 q}{m_{\rm pl} \epsilon}}\left(1 + \frac{12 - g^2_{\rm s}g^2_{\rm v}\alpha^2_{\rm ee}}{4}\frac{q}{k_{\rm TF}} + \dots \right)
\end{equation}
where we have introduced the density-dependent ``plasmon mass"
\begin{equation}
m_{\rm pl} = \frac{4 \pi n}{g_{\rm s} g_{\rm v}}~\frac{\hbar^2}{\varepsilon_{\rm F}}~,
\end{equation}
the fine structure constant $\alpha_{\rm ee} = e^2/(\hbar v \epsilon)$,  and the Thomas-Fermi screening vector $k_{\rm TF} = g_{\rm s} g_{\rm v} \alpha_{\rm ee} k_{\rm F}$. 

Note that $m_{\rm pl}$ tends to the bare electron mass in vacuum $m$ if we use the parabolic energy-momentum dispersion $\varepsilon_{\rm F} = \hbar^2 k^2_{\rm F}/ (2m)$ (while still using $g_{\rm s} g_{\rm v} = 4$). In graphene, however, $m_{\rm pl} = \hbar k_{\rm F}/v$, which (i) is $\propto \sqrt{n}$ and (ii) trivially depends on Planck's constant $\hbar$ because the Fermi energy in this material depends linearly on $\hbar$ (rather than quadratically). As explained in Ref.~\onlinecite{polini_pseudospin_2009}, because the MDF model Hamiltonian is not invariant under an ordinary Galileian boost, RPA is not exact for interacting systems of MDFs even in the limit $q \to 0$ (contrary to what happens in the conventional 2D  parabolic-band electron gas~\cite{Giuliani_and_Vignale}, where for $q \to 0$ the plasmon is protected from many-body renormalizations by Galileian invariance). When interactions between MDFs are treated beyond RPA the plasmon mass acquires a non-trivial density and coupling-constant dependence~\cite{polini_pseudospin_2009}. 

Last but not least, note that, at odds with what stated in Ref.~\onlinecite{hwang_prb_2007}, the first subleading correction to the RPA plasmon dispersion [second term in round brackets in Eq.~(\ref{eq:plasmon_dispersion})] {\it can} change sign: it is positive (like in the conventional 2D  parabolic-band electron gas~\cite{Giuliani_and_Vignale}) at weak coupling, but  becomes negative for $\alpha_{\rm ee} > \sqrt{3}/2$.

\subsection{Transverse channel}

In this Section we report explicit expressions for the transverse response 
$\chi^{(0)}_{\sigma_x\sigma_x}(q{\hat {\bm y}},\omega)$. 
As done in Sect.~\ref{sect:long_doped}, we introduce the contribution due to doping, $\delta \chi^{(0)}_{\sigma_x\sigma_x}(q{\hat {\bm y}},\omega)$, according to
\begin{equation}
\chi^{(0)}_{\sigma_x\sigma_x}(q{\hat {\bm y}},\omega) = \chi^{(0{\rm u})}_{\sigma_x\sigma_x}(q{\hat {\bm y}},\omega) + \delta \chi^{(0)}_{\sigma_x\sigma_x}(q{\hat {\bm y}},\omega)~.
\end{equation}
In what follows, we report only expressions for $\delta \chi^{(0)}_{\sigma_x\sigma_x}(q{\hat {\bm y}},\omega)$.

Similarly to what done in Sect.~\ref{sect:long_doped}, we introduce the complex function
\begin{equation}
G_{\rm T} (z) = z\sqrt{z^2 - 1} + \ln(z + \sqrt{z^2 - 1})~.
\end{equation}
Using $G_{\rm T}(z)$ and the function $F_{\rm T}(q,\omega)$ introduced above in Eq.~(\ref{eq:trans_F}) we find
\begin{eqnarray}
\delta \chi^{(0)}_{\sigma_x\sigma_x}(q{\hat {\bm y}},\omega)&=& \frac{\omega^2}{v^2 q^2}\frac{\varepsilon_{\rm F}}{2\pi v^2} - F_{\rm T}(q, \omega) \{
G_{\rm T}(\omega_+)  \nonumber \\ 
&-& \Theta(\omega_- - 1) [G_{\rm T}(\omega_-) + i \pi] \nonumber \\
&-& \Theta(1 - \omega_-) G_{\rm T}(-\omega_-)\}~.
\end{eqnarray}

Once again, we provide below more explicit expressions for $\delta \chi^{(0)}_{\sigma_x\sigma_x}(q{\hat {\bm y}},\omega)$ in terms of real functions. 

\subsubsection{Region A.1}
For 
%$\omega < vq$ and 
$\omega < vq - 2\varepsilon_{\rm F}$:
\begin{equation}
\Re e~\delta \chi^{(0)}_{\sigma_x\sigma_x} (q{\hat {\bm y}}, \omega) =  
\frac{\omega^2}{v^2 q^2}\frac{\varepsilon_{\rm F}}{2\pi v^2} - \frac{\sqrt{v^2 q^2 - \omega^2}}{16\pi v^2} a_{\rm T}(q,\omega)
\end{equation}
and $\Im m~\delta \chi^{(0)}_{\sigma_x\sigma_x} (q\hat{\bm{y}}, \omega) = 0$. Here 
\begin{eqnarray}
a_{\rm T}(q,\omega) &=& \arcsin\left(\omega_+\right) - \omega_+\sqrt{1 -\omega_+^2}\nonumber\\
&+& \arcsin\left(\omega_-\right) \nonumber - \omega_-\sqrt{1 - \omega_-^2}~.
\end{eqnarray}

\subsubsection{Region A.2}
For $\omega < vq$ and $\omega >|2\varepsilon_{\rm F}- vq|$:
\begin{eqnarray}
\Re e~\delta \chi^{(0)}_{\sigma_x\sigma_x} (q\hat{\bm{y}}, \omega) &=& - \frac{\sqrt{v^2q^2 - \omega^2}}{16 v^2} + \frac{\omega^2}{v^2 q^2}\frac{\varepsilon_{\rm F}}{2\pi v^2} \nonumber\\
&+& \frac{\sqrt{v^2 q^2 - \omega^2}}{16\pi v^2} b_{\rm T}(q,\omega)
\end{eqnarray}
and
\begin{eqnarray}
\Im m~\delta \chi^{(0)}_{\sigma_x\sigma_x} (q\hat{\bm{y}}, \omega) = - \frac{\sqrt{v^2 q^2 - \omega^2}}{16\pi v^2} c_{\rm T}(q,\omega)~.
\end{eqnarray}
Here
\begin{equation}
b_{\rm T}(q,\omega) = \arccos(\omega_-) + \omega_-\sqrt{1 - \omega_-^2}
\end{equation}
and
\begin{equation}
c_{\rm T}(q,\omega) = \ln\left(\omega_+ + \sqrt{\omega_+^2 - 1}\right) + \omega_+\sqrt{\omega_+^2 - 1}~.
\end{equation}

\subsubsection{Region A.3}
For $\omega < vq$ and $\omega<2\varepsilon_{\rm F}-vq$:
\begin{equation}
\Re e~\delta \chi^{(0)}_{\sigma_x\sigma_x} (q\hat{\bm{y}}, \omega) = - \frac{\sqrt{v^2q^2 - \omega^2}}{16 v^2} + \frac{\omega^2}{v^2 q^2}\frac{\varepsilon_{\rm F}}{2\pi v^2}
\end{equation}
and
\begin{eqnarray}
\Im m~\delta \chi^{(0)}_{\sigma_x\sigma_x} (q\hat{\bm{y}}, \omega) &=& - \frac{\sqrt{v^2 q^2 - \omega^2}}{16\pi v^2} d_{\rm T}(q, \omega)~.
\end{eqnarray}
Here
\begin{eqnarray}
d_{\rm T}(q,\omega) &=& \ln\left(\frac{\omega_+ + \sqrt{\omega_+^2 - 1}}{\omega_- + \sqrt{\omega_-^2 - 1}}\right) + \omega_+\sqrt{\omega_+^2 - 1} \nonumber\\
&-& \omega_-\sqrt{\omega_-^2 - 1}~.
\end{eqnarray}

\subsubsection{Region B.1}
For 
%$\omega > vq$ and 
$\omega > 2\varepsilon_{\rm F} + vq$:
\begin{equation}
\Re e~\delta \chi^{(0)}_{\sigma_x\sigma_x} (q\hat{\bm{y}}, \omega) = \frac{\omega^2}{v^2 q^2}\frac{\varepsilon_{\rm F}}{2\pi v^2} - \frac{\sqrt{\omega^2 - v^2 q^2}}{16\pi v^2} e_{\rm T}(q,\omega)
\end{equation}
and $\Im m~\delta \chi^{(0)}_{\sigma_x\sigma_x} (q\hat{\bm{y}}, \omega) = 0$. Here
\begin{eqnarray}
e_{\rm T} (q,\omega) &=& \ln\left(\frac{\omega_+ + \sqrt{\omega_+^2 - 1}}{\sqrt{\omega_-^2 - 1}-\omega_-}\right) + \omega_+\sqrt{\omega_+^2 - 1} \nonumber\\
&+& \omega_-\sqrt{\omega_-^2 - 1}~.
\end{eqnarray}

\subsubsection{Region B.2}
For  $\omega > vq$, $\omega > 2\varepsilon_{\rm F}-vq$, and $\omega < 2\varepsilon_{\rm F} +vq$:
\begin{equation}
\Re e~\delta \chi^{(0)}_{\sigma_x\sigma_x} (q\hat{\bm{y}}, \omega) = \frac{\omega^2}{v^2 q^2}\frac{\varepsilon_{\rm F}}{2\pi v^2} - \frac{\sqrt{\omega^2 - v^2 q^2}}{16\pi v^2} f_{\rm T}(q,\omega)
\end{equation}
and
\begin{equation}
\Im m~\delta \chi^{(0)}_{\sigma_x\sigma_x} (q\hat{\bm{y}}, \omega) = \frac{\sqrt{\omega^2 - v^2 q^2}}{16 v^2} -\frac{\sqrt{\omega^2 - v^2 q^2}}{16\pi v^2} g_{\rm T}(q,\omega) ~.
\end{equation}
Here
\begin{equation}
f_{\rm T} (q,\omega) = \ln\left(\omega_+ + \sqrt{\omega_+^2 - 1}\right) + \omega_+\sqrt{\omega_+^2 - 1}
\end{equation}
and
\begin{equation}
g_{\rm T} (q,\omega) = \arccos\left(\omega_-\right) + \omega_-\sqrt{1 - \omega_-^2}~.
\end{equation}

\subsubsection{Region B.3}
For $\omega> vq$ and $\omega <2\varepsilon_{\rm F} - vq$:
\begin{equation}
\nonumber \Re e~\delta \chi^{(0)}_{\sigma_x\sigma_x} (q\hat{\bm{y}}, \omega) = \frac{\omega^2}{v^2 q^2}\frac{\varepsilon_{\rm F}}{2\pi v^2} - \frac{\sqrt{\omega^2 - v^2 q^2}}{16\pi v^2} h_{\rm T} (q,\omega)
\end{equation}
and
\begin{equation}
\Im m~\delta \chi^{(0)}_{\sigma_x, \sigma_x} (q\hat{\bm{y}}, \omega) = \frac{\sqrt{\omega^2 - v^2 q^2}}{16 v^2}~.
\end{equation}
Here
\begin{eqnarray}
h_{\rm T} (q,\omega) &=& \ln\left(\frac{\omega_+ + \sqrt{\omega_+^2 - 1}}{\omega_- + \sqrt{\omega_-^2 - 1}}\right) + \omega_+\sqrt{\omega_+^2 - 1} \nonumber\\
&-&  \omega_-\sqrt{\omega_-^2 - 1} ~.
\end{eqnarray}

\begin{figure}[t]
\begin{center}
\tabcolsep=0cm
\begin{tabular}{cc}
\includegraphics[width=0.50\linewidth]{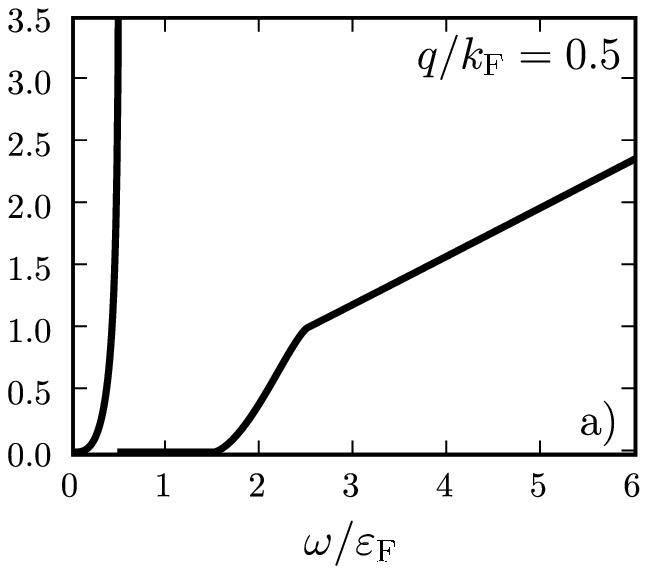}&
\includegraphics[width=0.50\linewidth]{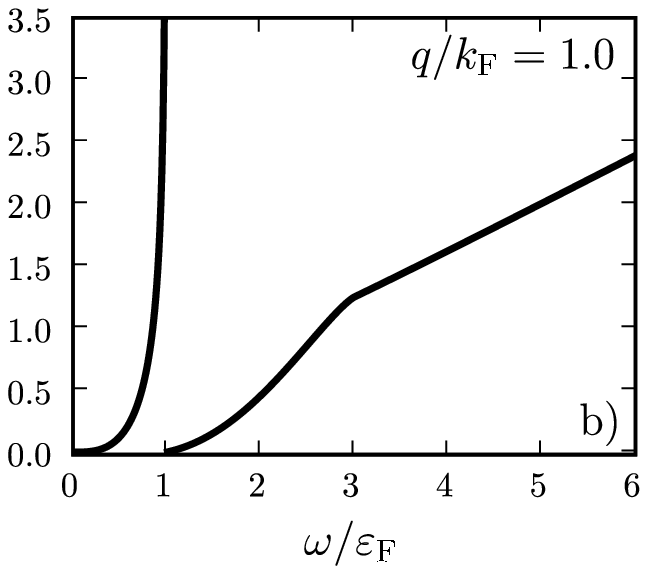}\\
\includegraphics[width=0.50\linewidth]{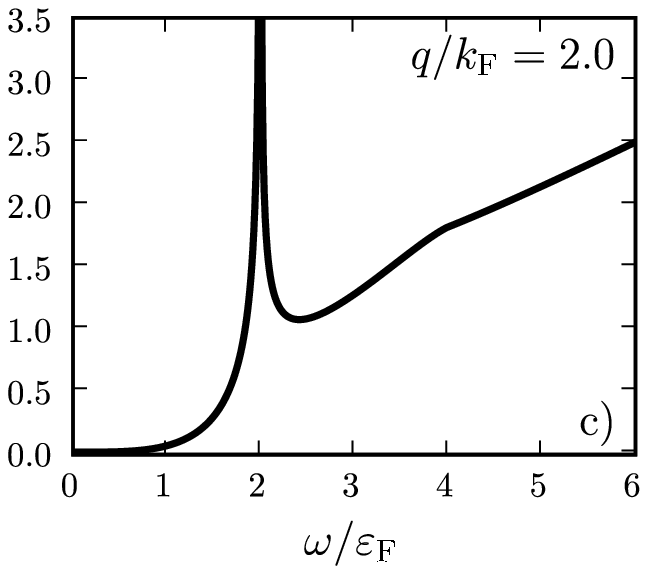}&
\includegraphics[width=0.50\linewidth]{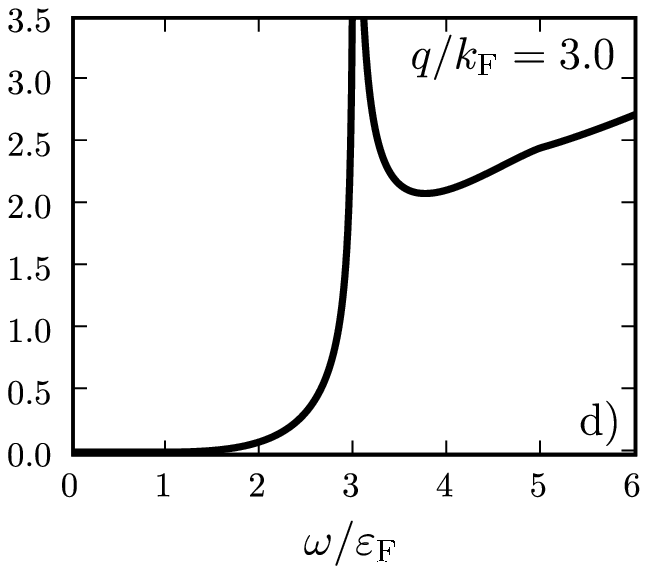}
\end{tabular}
\end{center}
\caption{The imaginary part of the longitudinal pseudospin response function, 
$-\Im m~\chi^{(0)}_{\sigma^x\sigma^x}({\bm q}=q {\hat x}, \omega)$ [in units of the density-of-states at the Fermi level, 
$\nu(0) = \varepsilon_{\rm F}/(2\pi v^2)$], as a function of $\omega/\varepsilon_{\rm F}$. 
a) $q=0.5~k_{\rm F}$, b) $q=1.0~k_{\rm F}$, c) $q=2.0~k_{\rm F}$, and d) $q=3.0~k_{\rm F}$. Singularities at $\omega = v q$ are clearly visible~\cite{polini_prb_2008}.\label{fig:two}}
\end{figure}

\begin{figure}[t]
\begin{center}
\tabcolsep=0cm
\begin{tabular}{cc}
\includegraphics[width=0.50\linewidth]{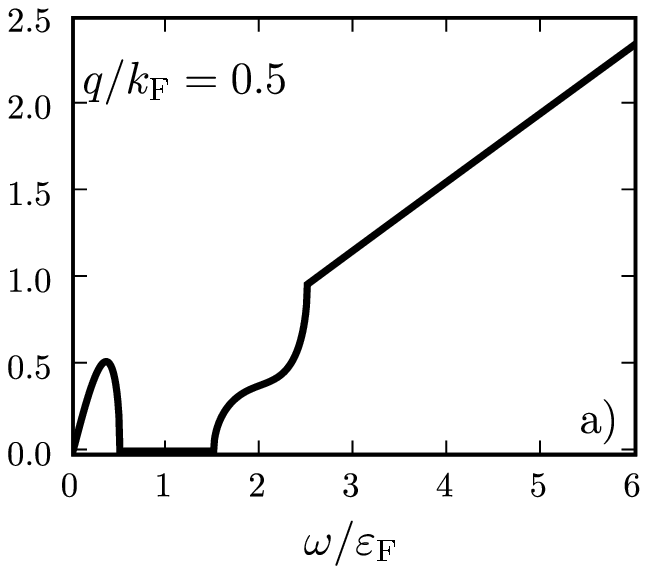}&
\includegraphics[width=0.50\linewidth]{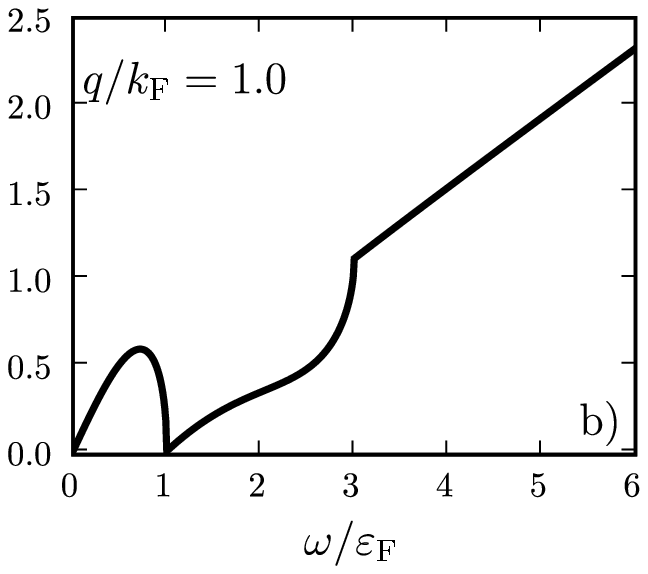}\\
\includegraphics[width=0.50\linewidth]{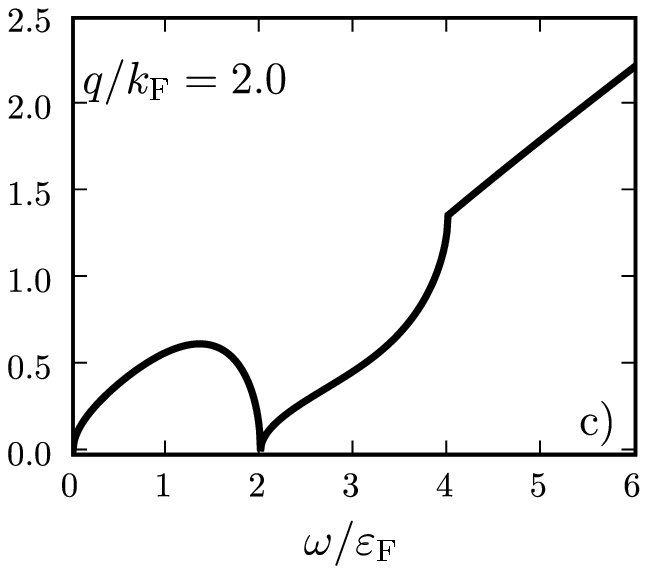}&
\includegraphics[width=0.50\linewidth]{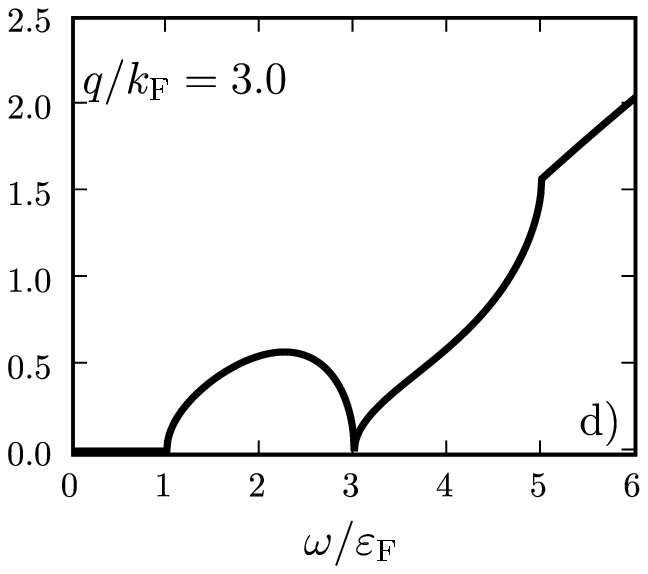}
\end{tabular}
\end{center}
\caption{The imaginary part of the transverse pseudospin response function, 
$-\Im m~\chi^{(0)}_{\sigma^x\sigma^x}({\bm q}=q {\hat y}, \omega)$ [in units of $\nu(0)$], as a function of $\omega/\varepsilon_{\rm F}$. 
a) $q=0.5~k_{\rm F}$, b) $q=1.0~k_{\rm F}$, c) $q=2.0~k_{\rm F}$, and d) $q=3.0~k_{\rm F}$.\label{fig:three}}
\end{figure}

\begin{figure}[t]
\begin{center}
\tabcolsep=0cm
\begin{tabular}{cc}
\includegraphics[width=0.50\linewidth]{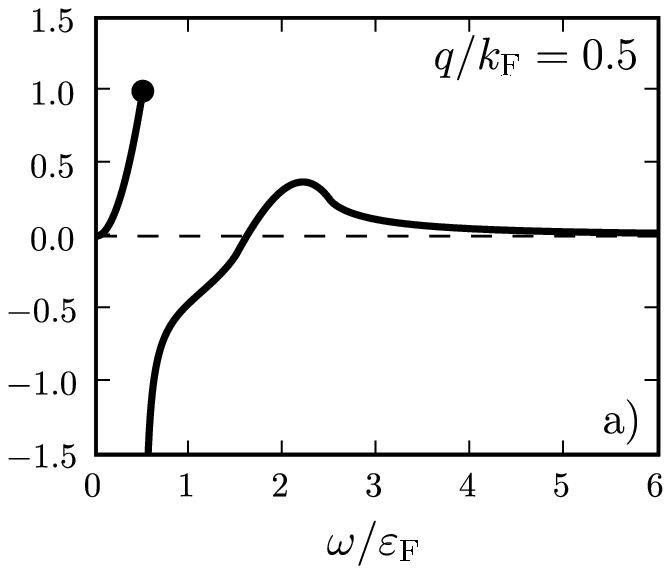}&
\includegraphics[width=0.50\linewidth]{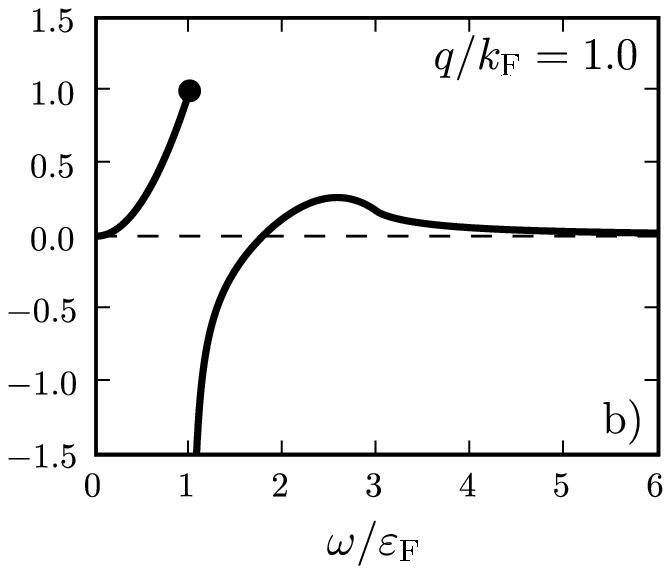}\\
\includegraphics[width=0.50\linewidth]{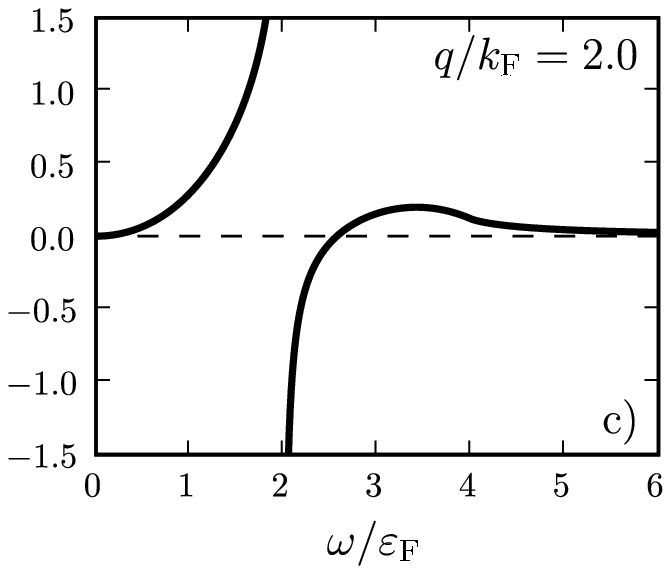}&
\includegraphics[width=0.50\linewidth]{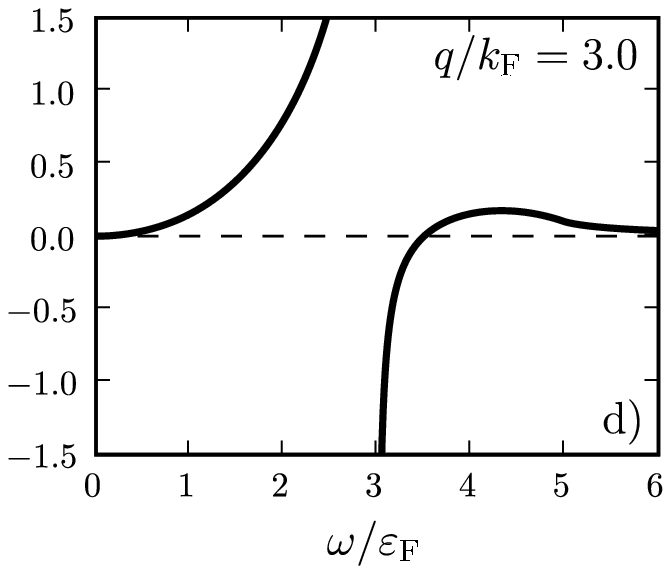}
\end{tabular}
\end{center}
\caption{The real part of the longitudinal pseudospin response function, 
$-\Re e~\chi^{(0)}_{\sigma^x\sigma^x}({\bm q}=q {\hat x}, \omega)$ [in units of $\nu(0)$], as a function of $\omega/\varepsilon_{\rm F}$. 
a) $q=0.5~k_{\rm F}$, b) $q=1.0~k_{\rm F}$, c) $q=2.0~k_{\rm F}$, and d) $q=3.0~k_{\rm F}$. The filled circles in panels a) and b) mean that the function $- \Re e~\chi^{(0)}_{\sigma^x\sigma^x}({\bm q}=q {\hat x}, \omega)$ takes exactly the value $1$ [in units of $\nu(0)$] at $\omega = vq$.\label{fig:four}}
\end{figure}

\begin{figure}[t]
\begin{center}
\tabcolsep=0cm
\begin{tabular}{cc}
\includegraphics[width=0.50\linewidth]{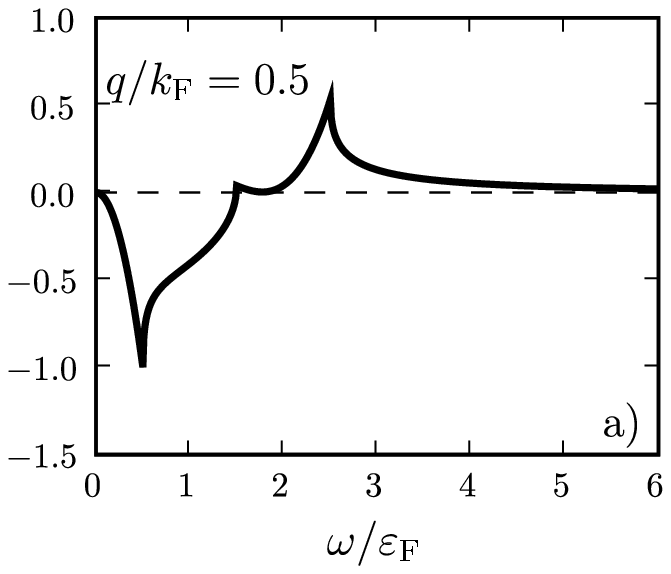}&
\includegraphics[width=0.50\linewidth]{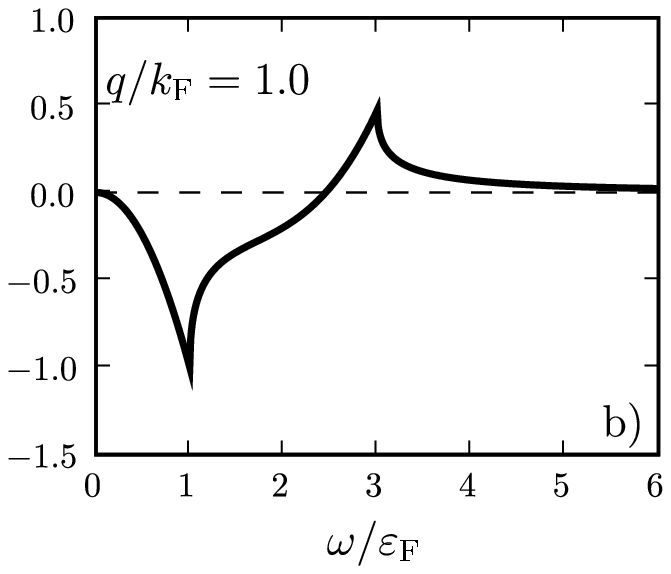}\\
\includegraphics[width=0.50\linewidth]{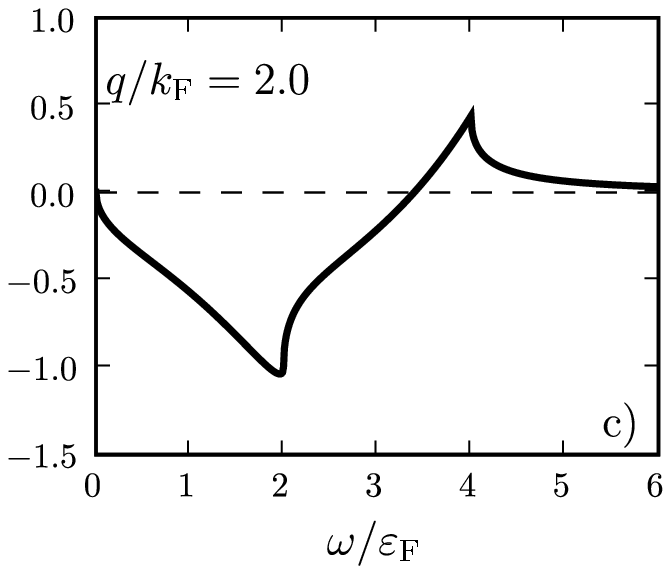}&
\includegraphics[width=0.50\linewidth]{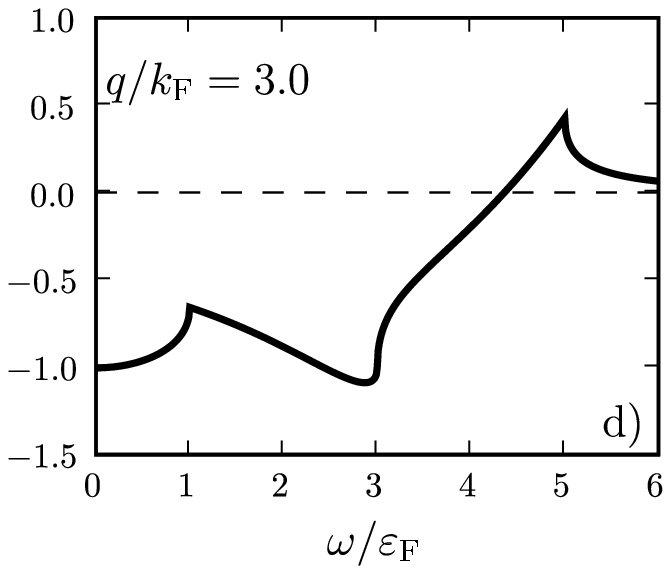}
\end{tabular}
\end{center}
\caption{The real part of the transverse pseudospin response function, 
$-\Re e~\chi^{(0)}_{\sigma^x\sigma^x}({\bm q}=q {\hat y}, \omega)$ [in units of $\nu(0)$], as a function of $\omega/\varepsilon_{\rm F}$. 
a) $q=0.5~k_{\rm F}$, b) $q=1.0~k_{\rm F}$, c) $q=2.0~k_{\rm F}$, and d) $q=3.0~k_{\rm F}$.\label{fig:five}}
\end{figure}

In Figs.~\ref{fig:two}-\ref{fig:five} we have reported plots of the longitudinal and transverse response functions. Note that the transverse response function goes always to zero at $\omega = vq$: in this case, indeed, the angle between ${\bm k}$ and ${\bm q}$ is either zero (intraband term) or $\pi$ (interband term). Recalling that in the transverse case ${\bm q}$ is directed along ${\hat {\bm y}}$, we thus find that either $\varphi_{{\bm k}+{\bm q}} = \varphi_{\bm k} = \pi/2$ (intraband term) or $\varphi_{{\bm k}+{\bm q}} = - \varphi_{\bm k} = \pi/2$ (interband term): in both cases the matrix element in Eq.~(\ref{eq:sum_angles}) vanishes. Finally, note that both 
$\Im m~\chi^{(0)}_{\sigma_x \sigma_x} (q\hat{\bm{x}}, \omega)$ and $\Im m~\chi^{(0)}_{\sigma_x \sigma_x} (q\hat{\bm{y}}, \omega)$ diverge linearly for $\omega \gg vq$.

\section{Diamagnetic susceptibility}
\label{sect:diamagnetic_suscept}

The transverse pseudospin response function allows us to calculate the orbital magnetization induced in doped graphene sheets by a static magnetic field. As discussed in detail in Sects.~3.4.3. and~4.5 of Ref.~\onlinecite{Giuliani_and_Vignale}, the noninteracting orbital magnetic susceptibility $\chi^{(0)}_{\rm orb}$ can be calculated from
\begin{equation}\label{eq:chiorb}
\chi^{(0)}_{\rm orb} = -\frac{v^2 e^2}{c^2} \lim_{q \to 0} \frac{\chi^{(0)}_{\sigma^x\sigma^x}(q {\hat {\bm y}}, 0)}{q^2}~.
\end{equation}
As stressed earlier in Sect.~\ref{sect:undoped}, before taking the limit in Eq.~(\ref{eq:chiorb}) the undoped contribution to the static response $\chi^{(0)}_{\sigma^x\sigma^x}(q {\hat {\bm y}}, 0)$ has to be regularized to restore gauge invariance by subtracting the cut-off dependent constant term $-\varepsilon_{\rm max}/(4\pi v^2)$. A simple inspection of the equations reported in the previous section allows us to write a compact expression for the static transverse response (as usual, per spin and valley):
\begin{eqnarray}\label{eq:static_transverse}
\chi^{(0)}_{\sigma_x \sigma_x} (q {\hat {\bm y}}, 0) &=& \frac{q}{16 v} \left\{ 1 - \frac{2}{\pi} \arcsin\left[\ell(q)\right]\right\} \nonumber \\
&+& \frac{\varepsilon_{\rm F}}{4\pi v^2} \sqrt{1 - \left(\frac{2k_{\rm F}}{q}\right)^2} \Theta\left(1 - \frac{2k_{\rm F}}{q}\right)\nonumber\\
\end{eqnarray}
where
\begin{equation}
\ell(q) = \frac{1}{2} \left(1 + \frac{2k_{\rm F}}{q}\right) - \frac{1}{2} \left|1 - \frac{2k_{\rm F}}{q}\right|~.
\end{equation}
We thus immediately see that for $\varepsilon_{\rm F} >0$ and for all wavevectors $q< 2 k_{\rm F}$,  $\ell(q)=1$ and $\chi^{(0)}_{\sigma_x \sigma_x} (q {\hat {\bm y}}, 0) \equiv 0$. This implies that the orbital magnetic susceptibility of noninteracting MDFs is {\it zero}. However, if $\varepsilon_{\rm F} =0$ $\chi^{(0{\rm u})}_{\sigma_x \sigma_x} (q {\hat {\bm y}}, 0) \propto q$ and thus the orbital magnetic susceptibility diverges in the undoped limit. More precisely, it is possible to show that in the limit $q \to 0$ the function $\chi^{(0)}_{\sigma_x \sigma_x} (q {\hat {\bm y}}, 0)/q^2$ is proportional to $\delta(\varepsilon_{\rm F})$. Indeed, if $\varphi(\varepsilon)$ is a test function (here $\varepsilon$ is a shorthand notation for $\varepsilon_{\rm F}$),
\begin{eqnarray}
{\cal I} &\equiv& \lim_{q \to 0} \int_{-\infty}^{+\infty} d\varepsilon~\frac{\chi_{\sigma_x \sigma_x}^{(0)} (q {\hat {\bm y}}, 0)}{q^2} \varphi(\varepsilon) \nonumber\\
&=&  2~\lim_{q \to 0} \int_0^{vq/2}d \varepsilon~\frac{1}{16 vq}~\Bigg[ 1 - \frac{2}{\pi} \arcsin\left(\frac{2 \varepsilon}{vq}\right) \nonumber\\ 
&+& \frac{2}{\pi}~\frac{2 \varepsilon}{vq} \sqrt{1 - \left(\frac{2 \varepsilon}{vq}\right)^2}\Bigg]\varphi(\varepsilon)~.
\end{eqnarray}
Introducing the dimensionless variable $x = 2 \varepsilon/ (vq)$ we find
\begin{eqnarray}
{\cal I} &= &  \lim_{q \to 0} \Bigg\{ \frac{1}{16} - \frac{1}{8 \pi} \Bigg[ \int_0^1 dx~\arcsin (x) \nonumber\\
&-& \int_0^1 dx~x \sqrt{1 - x^2 }\Bigg] \Bigg\}\varphi({\bar \varepsilon}) = \frac{1}{6\pi}~\varphi(0)~,
\end{eqnarray}
where ${\bar \varepsilon} \in [0, vq/2]$. In summary, introducing the $g_{\rm s}g_{\rm v} =4$ degeneracy factor, we find~\cite{footnote_finite_T} 
\begin{equation}\label{eq:orb_final} 
\chi^{(0)}_{\rm orb} = - \frac{g_{\rm s}g_{\rm v}}{6\pi} \frac{e^2 v^2}{c^2} \delta(\varepsilon_{\rm F})~,
\end{equation}
in perfect agreement with Ref.~\onlinecite{mcclure}. As explained by McClure~\cite{mcclure}, the origin of the large (infinite at $T=0$) diamagnetism in undoped graphene can be understood qualitatively. When the magnetic field $B$ is turned on, group of states, which were originally distributed in energy, coalesce into Landau levels. In undoped graphene, states which had negative energy coalesce into the $n=0$ Landau level, thus increasing the energy of the system, which will respond to the field with a negative orbital susceptibility. The total energy (per unit area) in the absence of the field of the group of electrons which condense into the $n=0$ Landau level is
\begin{eqnarray}\label{eq:energy_before}
E_0 &=& \int_{-\omega_{\rm c}/\sqrt{2}}^{+\omega_{\rm c}/\sqrt{2}} d\varepsilon~\varepsilon \nu(\varepsilon) f(\varepsilon) \nonumber\\
&=&  \int_{0}^{+\omega_{\rm c}/\sqrt{2}} d\varepsilon~\varepsilon \nu(\varepsilon)[f(\varepsilon)- f(-\varepsilon)]~,
\end{eqnarray}
where $\omega_{\rm c} = \sqrt{2} v/ \ell_{\rm B} \propto \sqrt{B}$ is the MDF cyclotron frequency [$\ell_B= \sqrt{c/(eB)}$ being the magnetic length], $\nu(\varepsilon) = |\varepsilon|/(2\pi v^2)$ is the density-of-states at $B=0$, and $f(\varepsilon)$ is the Fermi-Dirac distribution function. The extreme of integrations in the first line of Eq.~(\ref{eq:energy_before}) ensure that the number of states
\begin{eqnarray}
N_0 &=& \int_{-\lambda}^{+\lambda} d\varepsilon~\nu(\varepsilon) = \frac{\lambda^2}{2\pi v^2}
\end{eqnarray}
can be accommodated into the $n=0$ Landau level, which has degeneracy $e B / (2\pi c)$ per unit area, {\it i.e.} $\lambda = \omega_{\rm c}/\sqrt{2}$.

In the zero-temperature limit 
\begin{equation}
E_0 = - \frac{1}{2\pi v^2} \int_{0}^{+\omega_{\rm c}/\sqrt{2}} d\varepsilon~\varepsilon^2 = - \frac{1}{12\sqrt{2}\pi v^2}\omega^3_{\rm c} \propto - B^{3/2}~.
\end{equation}
The total energy in the presence of the field is zero (because turning on the $B$ field all the electrons condense into the $n=0$ Landau level, which is at zero energy). Thus the change in energy with the field is $\Delta E = - E_0 \propto B^{3/2}$. The susceptibility is thus
\begin{equation}
\chi^{(0)}_{\rm orb} = - \lim_{B \to 0} \frac{1}{B}~\frac{\partial (\Delta E)}{\partial B} \propto - \frac{1}{\sqrt{B}}~,
\end{equation}
which diverges for $B \to 0$.

Introducing the usual Bohr magneton, 
$\mu_{\rm B} = e\hbar/(2 m c)$, 
and restoring $\hbar$, the orbital magnetic susceptibility (\ref{eq:orb_final}) can be written as
\begin{equation}
\chi^{(0)}_{\rm orb} =  -\left(\frac{g \mu_{\rm B}}{2}\right)^2 \frac{2 g_{\rm s} g_{\rm v} m^2 v^2}{3\pi \hbar^2} \delta(\varepsilon_{\rm F})~,
\end{equation}
$g=2$ being the bare electron $g$-factor in vacuum. We recall that the Pauli spin susceptibility of 2D MDFs is
\begin{equation}
\chi^{(0)}_{\rm P} = \left(\frac{g \mu_{\rm B}}{2}\right)^2~\frac{g_{\rm s} g_{\rm v} \varepsilon_{\rm F}}{2 \pi \hbar^2 v^2}~.
\end{equation}
In the conventional 2D parabolic-band electron gas $\chi^{(0)}_{\rm orb} = -\chi^{(0)}_{\rm P}/3$. For 2D MDFs we see that the density dependence of $\chi^{(0)}_{\rm orb}$ is completely different from that of $\chi^{(0)}_{\rm P}$.

\section{Summary and conclusions}
\label{sect:summary}

In summary, we have calculated analytically the longitudinal and transverse pseudospin-pseudospin linear-response functions of noninteracting massless Dirac fermions. As expected, because of the continuity equation that relates the density operator with the longitudinal current operator, the longitudinal response function is determined by the density-density response function (apart from an anomalous commutator term, which has to be handled carefully). We have used the transverse pseudospin response function to calculate the orbital magnetization induced in doped graphene sheets by a static magnetic field, finding perfect agreement with earlier calculations~\cite{mcclure} based on the explicit use of the Landau level structure of two-dimensional massless Dirac fermions.

The results presented in this work constitute a very useful starting point for the construction of approximate 
response functions that would take into account electron-electron interactions. For example, within the random phase approximation~\cite{Giuliani_and_Vignale} 
the pseudospin-pseudospin response functions of the interacting system can be written as
\begin{equation}
\chi_{\sigma^x\sigma^x}({\bm q}, \omega)= \frac{\chi^{(0)}_{\sigma^x\sigma^x}({\bm q}, \omega)}{1-v_q 
\chi^{(0)}_{\sigma^x\sigma^x}({\bm q}, \omega)}~.
\end{equation}

Our results are also very useful in view of future applications of current-density-functional theory~\cite{vignale_prl_1987} 
to doped graphene sheets in the presence of time- and spatially-varying vector potentials (see also comments in Sect.~II of Ref.~\onlinecite{tomadin_prb_2008}).

\acknowledgments

M.P. was partially supported by the CNR-INFM ``Seed Projects". G.V. acknowledges support from NSF Grant No. DMR-0705460.

\appendix

\section{Details on the analytical calculation of the transverse undoped response function}
\label{app:undoped}

\subsection{Imaginary part}
\label{app:sub:im_undoped}

In the undoped limit and for $\omega>0$, the imaginary part of the transverse 
response function is given by 
\begin{eqnarray}\label{eq:appe_details}
\Im m~\chi^{(0{\rm u})}_{\sigma_x\sigma_x}(q{\hat {\bm y}},\omega) &=& - \pi \int\frac{d^2{\bm k}}{(2\pi)^2} 
\delta\left(\omega - vk - v|{\bm k} +{\bm q}|\right)\nonumber\\
& \times & \frac{1 - \cos(\varphi_{{\bm k} + {\bm q}} + \varphi_{\bm k})}{2}~.
\end{eqnarray}
This expression can be easily obtained from Eq.~(\ref{eq:lindhard}) by retaining only the interband term corresponding to 
$\lambda = +$ and $\lambda' = - $ (because $\omega>0$). The cosine term in the second line of Eq.~(\ref{eq:appe_details}) 
can be easily manipulated to give
\begin{equation}\label{eq:cos_sum}
\cos(\varphi_{{\bm k} + {\bm q}} + \varphi_{\bm k}) = \frac{k(1-2\sin^2\varphi_{\bm k}) - q\sin\varphi_{\bm k}}{|{\bm k} + {\bm q}|}~.
\end{equation}
For a given value of $\omega$ the delta function in Eq.~(\ref{eq:appe_details}) gives a non-zero contribution to the ${\bm k}$-integration if and only if 
\begin{equation}\label{eq:und_trans_conditions}
\omega - vk = v|{\bm k} + {\bm q}|~,
\end{equation}
which can be solved for $k$ yielding 
\begin{equation}
k = \frac{\omega^2 - v^2q^2}{2v(\omega + vq\sin\varphi_{\bm k})}~.
\end{equation}
Performing the integration over $k$ we are left with
\begin{eqnarray}
\Im m~\chi^{(0{\rm u})}_{\sigma_x\sigma_x}(q{\hat {\bm y}},\omega) &=& - \frac{\omega^2 -v^2q^2}{16\pi v^2} \Theta(\omega-vq) \nonumber\\
&\times& \int_0^{2\pi} d\varphi_{\bm k} \frac{(\omega\sin\varphi_{\bm k} + vq)^2}{(\omega + vq\sin\varphi_{\bm k})^3}~.\nonumber\\
\end{eqnarray}
The angular integration can be easily performed in the complex plane $z = \exp(i\varphi_{\bm k})$. We thus find
\begin{equation}
\int_0^{2\pi} d\varphi_{\bm k} \frac{(\sin\varphi_{\bm k} + vq/\omega)^2}{(\omega/vq + \sin\varphi_{\bm k})^3} 
= \oint_{\cal C} dz \frac{[(z - z_1) (z - z_2)]^2}{[(z - z_{+}) (z - z_{-})]^3}~,\nonumber\\
\end{equation}
where ${\cal C}$ is the unit-radius circle in the complex plane and
\begin{eqnarray}
z_{1, 2} &=& \frac{-ivq \pm \sqrt{\omega^2 - v^2q^2}}{\omega}~,\\
z_{\pm} &=& i\frac{-\omega \pm \sqrt{\omega^2 - v^2q^2}}{vq} ~.
\end{eqnarray}
It is easy to see that $z_{+}z_{-} = 1$ and that $|z_{+}| < 1$, which implies $|z_{-}| > 1$. 
Calculating the residue in $z_{+}$ (which is a third-order pole) we finally find Eq.~(\ref{eq:im_trans_undoped}).

\subsection{Real part}
\label{app:sub:re_undoped}

We now pass to calculate $\Re e~\chi^{(0{\rm  u})}_{\sigma_x\sigma_x} (q{\hat {\bm y}}, \omega)$ by using the Kramers-Kr\"onig dispersion relation, {\it i.e.}
\begin{eqnarray}
& & \Re e~\chi^{(0{\rm  u})}_{\sigma_x\sigma_x} (q{\hat {\bm y}}, \omega) = \frac{2}{\pi} {\cal P} \int_0^{\infty}\omega'd\omega'\frac{\Im m~\chi^{(0{\rm u})}_{\sigma_x\sigma_x}(q{\hat {\bm x}},\omega')}{\omega'^2 - \omega^2}\nonumber\\
& = & -\frac{1}{8\pi v^2}\lim_{\omega_{\rm max} \to \infty} {\cal P} \int_{vq}^{\omega_{\rm max}}\omega'd\omega'\frac{\sqrt{\omega'^2-v^2q^2}}{\omega'^2 - \omega^2} .
\end{eqnarray}
Performing the change of variable $t^2 = \omega'^2 - v^2q^2$ and defining $t_{\rm max} = \sqrt{\omega_{\rm max}^2 - v^2q^2}$ we find
\begin{eqnarray}\label{eq:unf_long}
\Re e~\chi^{(0{\rm  u})}_{\sigma_x\sigma_x} (q{\hat {\bm y}}, \omega) &=& -\frac{t_{\rm max}}{8\pi v^2} 
- \frac{\omega^2- v^2q^2}{8\pi v^2}\nonumber\\
&\times& {\cal P}\int_{0}^{\infty}\frac{dt}{t^2 - (\omega^2 - v^2q^2)}~,
\end{eqnarray}
where the first term on the r.h.s. diverges in the limit $\omega_{\rm max} \to \infty$. Considering that $\omega_{\rm max}$ is a maximum {\it excitation} 
energy [and thus it is twice the cut-off $\varepsilon_{\rm max}$ introduced after Eq.~(\ref{eq:re_long_undoped})], 
this term can be re-written as 
\begin{equation}
-\frac{t_{\rm max}}{8\pi v^2} = -\frac{\varepsilon_{\rm max}}{4\pi v^2}\sqrt{1-\left(\frac{vq}{2\varepsilon_{\rm max}}\right)^2}~.
\end{equation}
In the limit $\varepsilon_{\rm max} \to \infty$ this terms thus tends to $-\varepsilon_{\rm max}/(4\pi v^2)$. The second integral gives
\begin{eqnarray}
{\cal P}\int_{0}^{\infty}\frac{dt}{t^2 - (\omega^2 - v^2q^2)} = \frac{\pi}{2}\frac{\Theta(vq - \omega)}{\sqrt{v^2q^2 - \omega^2}}~.
\end{eqnarray}
Summing together these two contributions we find immediately the final result (\ref{eq:trans}).

\section{Details on the analytical calculation of the transverse doped response function}
\label{app:doped}

In this Appendix we report some details on the analytical calculation of the transverse doped response function. In the doped case it turns out more convenient to perform the calculation on the imaginary frequency axis, {\it i.e.} $\omega \rightarrow i\omega$ in Eq.~(\ref{eq:lindhard}):
\begin{eqnarray}
\chi^{(0)}_{\sigma_x\sigma_x} (q\hat{{\bm y}}, \omega) &=& \frac{1}{S}\sum_{{\bm k}, \lambda, \lambda'} \frac{n^{(0)}_{{\bm k}, \lambda'} - n^{(0)}_{{\bm k} + {\bm q}, \lambda}}{i \omega + \lambda'vk - \lambda v|{\bm k} + {\bm q}|} \nonumber\\
&\times& \frac{1 + \lambda\lambda'\cos\left(\varphi_{{\bm k} + {\bm q}} + \varphi_{\bm k}\right)}{2}~.
\end{eqnarray}
Subtracting the undoped contribution and doing simple algebraic manipulations we find
\begin{equation}\label{eq:cc}
\delta\chi^{(0)}_{\sigma_x\sigma_x} (q\hat{{\bm y}}, \omega) = J(q, \omega) + J^*(q, \omega)~,
\end{equation}
where
\begin{equation} \label{eq:J_integral}
J(q, \omega) = \frac{1}{4 \pi^2} \int_{0}^{k_{\rm F}} k dk \int_{0}^{2\pi} d\varphi_{\bm k} f(\bm k, q, \omega)
\end{equation}
with
\begin{equation}
f({\bm k}, q, \omega) = \frac{2v k \sin^2\varphi_{\bm k} + vq\sin\varphi_{\bm k} - (i \omega + 2vk)}{2 v^2 k q \sin\varphi_{\bm k} + \omega^2  + v^2q^2 - 2i\omega v k}~.
\end{equation}
Once again the angular integration can be easily performed in the complex plane $z = \exp{(i \varphi_{\bm k})}$:
\begin{eqnarray}
I(q,\omega) &=& \frac{1}{4 \pi^2} \int_{0}^{2\pi} d\varphi_{\bm k} f(\bm k, q, \omega) \nonumber \\
&=& - \frac{1}{8 \pi^2 v^2 k q } \oint_{\cal C} dz
\frac{P(z)}{z^2(z - z_+)(z - z_-)} ~, \nonumber\\
\end{eqnarray}
with $P(z) = v k (z^2 - 1)^2 + i v q z (z^2 - 1) + 2 (i \omega + 2vk)z^2$,
\begin{eqnarray}
z_{\pm} &=& \frac{-i[v^2q^2 - (i\omega)^2  - 2i\omega v k]}{2v^2 k q} \nonumber\\
&\pm& \frac{\sqrt{v^2q^2 - (i\omega)^2}\sqrt{(i\omega + 2vk)^2 - v^2q^2}}{2v^2 k q} ~.
\end{eqnarray}
It is possible to show that $z_+ z_- = -1$ and $|z_+| < 1$. Calculating the residues 
in $z = 0$ (a second-order pole) and in $z = z_+$ (a first-order pole) and performing the integration over $k$ in Eq.~(\ref{eq:J_integral}) we finally find
\begin{eqnarray}
J(q, \omega) &=& -\frac{\sqrt{v^2 q^2 -(i\omega)^2}}{16\pi v^2} \left[\arcsin(\Omega)
+  i\Omega\sqrt{\Omega^2 - 1}\right] \nonumber\\
&+& \frac{\varepsilon_{\rm F}}{4\pi v^2}\frac{(i\omega)^2}{v^2 q^2}
\end{eqnarray}
where $\Omega = (2\varepsilon_{\rm F} + i\omega)/(vq)$. 
Using this result in Eq.~(\ref{eq:cc}) and performing a standard procedure of analytical continuation one finds 
the results in Sect.~\ref{sect:doped}.


\begin{thebibliography}{77}
\bibitem{footnote_finite_T}
	At finite temperature Eq.~(\ref{eq:orb_final}) reads
	$$
	\chi^{(0)}_{\rm orb} = - \frac{g_{\rm s}g_{\rm v}}{24 \pi} \frac{e^2 v^2}{c^2} 
	\frac{1}{k_{\rm B} T} {\rm sech}^2 \left(\frac{\mu}{2 k_{\rm B} T}\right)~,
	$$
	where $\mu$ is the chemical potential.
\end{thebibliography}
\end{document}